\documentclass[10pt]{article}
\usepackage{amsmath}
\usepackage{amsthm}
\usepackage{amsfonts}
\usepackage{amssymb}
\usepackage{url}
\usepackage{eucal}
\usepackage{mathrsfs}
\usepackage{float}
\usepackage[german,english]{babel}
\usepackage[T1]{fontenc}%
\usepackage{url}
\usepackage{epsfig}
\usepackage{epstopdf}
\usepackage{xcolor}

\newcommand\fig[1] {{\rm Figure}~\ref{fig:#1}}

\newcommand\labfig[1] {\label{fig:#1}}

\newcommand{\bfm}[1]{\mbox{\boldmath ${#1}$}}

\newcommand\eq[1] {(\ref{#1})}
\newcommand{\beqa}{\begin{eqnarray}}
\newcommand{\eeqa}[1]{\label{#1}\end{eqnarray}}
\newcommand{\beq}{\begin{equation}}
\newcommand{\eeq}[1]{\label{#1}\end{equation}}

\newcommand{\Grad}{\nabla}
\newcommand{\Div}{\nabla \cdot}

\newcommand{\Md}{\partial}

\newcommand{\Ga}{\alpha}
\newcommand{\Gb}{\beta}
\newcommand{\Gd}{\delta}

\newcommand{\Gg}{\gamma}
\newcommand{\Gc}{\chi}

\newcommand{\Go}{\omega}

\newcommand{\BGs}{\bfm\sigma}

\newcommand{\bpm}{\begin{pmatrix}}
	\newcommand{\epm}{\end{pmatrix}}

\def\b0{\bf 0}

\def\Be{{\bf e}}

\def\Bj{{\bf j}}
\def\Bk{{\bf k}}

\def\Bx{{\bf x}}

\begin{document}
\vspace{-1in}
\title{Field patterns: a new type of wave with infinitely degenerate band structure}
\author{Ornella Mattei and Graeme W. Milton\\
	\small{Department of Mathematics, University of Utah, Salt Lake City UT 84112, USA}}
\date{}
\maketitle
\begin{abstract}
	\noindent
Field pattern materials (FP--materials) are space--time composites with PT--symmetry in which the one--dimensional--spatial distribution of the constituents changes in time in such a special manner to give rise to a new type of waves, which we call field pattern waves (FP--waves) [G. W. Milton and O. Mattei, Proc. R. Soc. A 473, 20160819 (2017); O. Mattei and G. W. Milton, https://arxiv.org/abs/1705.00539 (2017)]. 
Specifically, due to the special periodic space--time geometry of these materials, when an instantaneous disturbance propagates through the system, the branching of the 
characteristic lines at the space--time interfaces between phases does not lead to a chaotic cascade of disturbances but concentrates on an orderly pattern of disturbances: this is the field pattern.  
By applying Bloch--Floquet theory we find that the dispersion diagrams associated with these FP--materials are infinitely degenerate: associated with each point
on the dispersion diagram is an infinite space of Bloch functions, a basis for which are generalized functions each concentrated on a field pattern, paramaterized by a variable
that we call the launch parameter. The dynamics separates into independent dynamics on the different field patterns, each with the same dispersion relation.  
\end{abstract}



\section{Introduction}

Waves in one-dimensional linear systems have a long history, going back to the study of Vincenzo Galilei with his son Galileo Galilei of 
vibrational waves on a string and in columns of air. Interesting effects occur when the moduli are inhomogeneous,
the most dramatic being the occurence in spatially periodic media of frequency bands where waves do not propagate,
and wave localization in spatially disordered media. One would think that there is nothing strikingly new to be found in
the study of waves in one-dimensional systems with varying moduli. However we find that there are some spectacularly
novel features associated with waves in certain one-dimensional linear media with moduli that vary both in
space and in time. One feature, remarked on in \cite{Mattei:2017:FPW}, is the appearance of a new type of wave generated from an instantaneous
source whose amplitude, unlike a conventional wake, does not go to zero away from the wave-front. Here we show
that the associated band structure in the dispersion diagram is infinitely degenerate. Each point on the dispersion
diagram, with a crystal wavenumber $k$ and crystal frequency $\Go(k)$, is associated with infinitely many eigenfunctions, each a generalized function concentrated on a discrete pattern
known as a field pattern, parameterized by a variable $\phi$ reflecting
the geometry of the field pattern that we call the launch parameter. 
The dynamics separates into dynamics on the infinitely many individual field patterns, each characterized by the same
dispersion relation. This infinite degeneracy is radically different to the degeneracy in atomic, photonic, phononic, or platonic crystals,
where at each crystal wavevector $\Bk$ the eigenspace associated with an eigenvalue $\Go(\Bk)$ in the discrete spectrum is finite dimensional, with the
dimension often reflecting the symmetries, leaving $\Bk$ invariant, that the system has relative to those of each eigenfunction.

We remark that a trivial example of infinitely degenerate bandstructure is obtained by taking a homogeneous material in space-time with one spatial
dimension with dispersion relation $\Go=\pm k$ and choose to regard it as being periodic in space and time with a square unit cell. Then band folding in $k$
followed by band folding in $\Go$ leads to infinite degeneracy: the Bloch functions can be taken as the plane waves $e^{k(x-t)}e^{n(x-t)}$ and $e^{k(x+t)}e^{m(x+t)}$ 
where $m$ and $n$ are integers.

%
The periodic systems in which this is achieved are field
pattern materials (henceforth, FP--materials) \cite{Milton:2017:FP,Mattei:2017:FPW}. These are special PT--symmetric (see, e.g., \cite{Bender:1998:SSD})
space--time microstructures in one spatial dimension plus time (P denotes parity symmetry, symmetry with respect to the spatial coordinate, and T denotes time symmetry) 
for which the special geometry of the constituent materials is such that, when a disturbance propagates through the system, the interaction disturbance--microstructure gives rise to a periodic space--time pattern of disturbances called, indeed, a field pattern. In other words, field pattern materials are simply one-dimensional media with time--dependent properties in which wave propagation exhibits brand--new features. PT--symmetry plays a crucial role in FP--materials (see \cite{Mattei:2017:FPW}): if the symmetry is broken then field patterns support both propagating waves and also waves that can blow-up exponentially with time, or decay exponentially
with time, whereas if the symmetry is unbroken, all the modes are propagating modes, with no blow-up.


Dynamic materials, first studied in the Fifties (e.g., \cite{Cullen:1958:TWP,Morgenthaler:1958:VME}), are raising increasing interest due to their exotic wave propagation properties, especially with reference to time modulation of the material parameters of photonic \cite{Joannopoulos:2008:ISS} and phononic crystals \cite{Deymier:2013:AMP}. Examples of such interesting phenomena are: the creation of effective magnetic fields for photons \cite{Fang:2012:REM};
the frequency splitting (analogous to Brillouin scattering) due to the interaction between an incident harmonic longitudinal wave and a
time-dependent phononic crystal \cite{Croenne:2017:BSL}; the transistor-like behavior of a helicoidal phononic crystal \cite{Li:2014:WTI}; and the breaking of time--reversal symmetry in time-modulated phononic crystals to realize one--way propagation devices \cite{Nassar:2017:MPC}, just to name a few. Very interestingly, a special type of space--time microstructure has been created called a ``discrete time crystal'' \cite{Zhang:2017:ODT,Choi:2017:ODT}: in the same way ``ordinary crystals'' form due to the breaking of spatial translational symmetry, discrete time crystals form due to the breaking of the time translational symmetry.

\section{Field pattern materials}

Here we briefly report the main features of FP--materials and we refer the reader to the papers \cite{Milton:2017:FP,Mattei:2017:FPW} for further details. For simplicity, as in those papers, we focus on FP--materials with one spatial dimension
and one time dimension. As there, we find it convenient to think of the wave--equation
with space and time-dependent moduli as being equivalent to a ``conductivity'' (or dielectric) equation with a real-valued
``conductivity tensor field'' having positive moduli in the spatial direction and negative moduli in the time direction. 
This allows us to use the language of the conductivity problem, but we stress that we are analyzing the one dimensional wave equation.
Thus,
in a one-spatial-dimension-plus-time $p$-component space-time composite, the ``conductivity'' tensor field takes the form
\beq \BGs(\Bx)=\sum_{i=1,p}\Gc_i(\Bx)\BGs_i, \eeq{1.2}
where $\Bx=(x,t)$, $\BGs_i$ is the conductivity tensor of phase $i$,, with $i=1,\dots,p$, and $\Gc_i(\Bx)$ is the characteristic function of phase $i$, equal to 1 if $\Bx$ belongs to phase $i$ and zero otherwise.
We suppose that all the $\BGs_i$, with $i=1,\dots,p$, take the form
\begin{equation}\label{1.3}
\BGs_i=
\begin{pmatrix} 
\Ga_i & 0 \\
0 & -\Gb_i 
\end{pmatrix}
\end{equation}
where $\Ga_i$ and $-\Gb_i$, with $\Ga_i$ and $\Gb_i$ both being positive, represent the ``conductivities'' in the space and time directions.

The dynamics is of course determined by the Green function. To find it one needs to solve for ``the electric current'' $\Bj(\Bx)$, 
the ``electric field'' $\Be(\Bx)$, and the ``electric potential'' $V(\Bx)$ satisfying the following set of equations
\beq \Bj(\Bx)=\BGs(\Bx)\Be(\Bx),\quad \Div\Bj=0, \quad \Be=-\Grad V, \eeq{1.1}
and the following initial conditions (assuming the medium to be infinite in the $x$ direction):
\beq V(x,0)=v_0H(x-a), \quad j_t(x,0)=j_0\Gd(x-a), \eeq{1.7a}
with $H(y)$ the Heaviside function, $\Gd(y)$ the Dirac delta function, and $v_0$ and $j_0$ constants. Thus, at time $t=0$, we are injecting, in the time direction ($j_t$ is the $t$-component of $\Bj$), a total current flux $j_0$ concentrated at $x=a$, and correspondingly there is a jump in potential.

Note that the global potential $V(\Bx)$ can be written as 
\beq V(\Bx)=\sum_{i=1,p}\Gc_i(\Bx)V_i(\Bx), \eeq{V_global}
$V_i(\Bx)$ being the potential in phase $i$, $i=1,\dots,p$, fulfilling within that phase the following wave equation, obtained by combining equations \eq{1.1} and \eq{1.3}:
\beq \Ga_i\frac{\Md^2 V_i}{\Md x^2}= \Gb_i\frac{\Md^2 V_i}{\Md t^2}\,. \eeq{1.4}
The D'Alembert solution \eq{1.4} gives the local (not global) solution in phase $i$ as the sum of two independent waves:
\beq V_i(x,t)=V^+_i(x-c_it)+V^-_i(x+c_it) \eeq{1.5}
with $V^+_i(x-c_it)$ the wave moving upwards to the right in a space--time diagram and $V^-_i(x+c_it)$ the wave moving upwards to the left, with the wave speed $c_i$ defined as
\beq c_i=\sqrt{\Ga_i/\Gb_i}. \eeq{1.6}
Due to the initial conditions \eq{1.7a}, $V_i(x,t)$ is piecewise constant in phase $i$: the potential jumps occur across characteristic lines and are given by the product of the modulus of the current flowing through that characteristic and an impedance coefficient $\Gg_i$, given by 
\beq \Gg_i=\frac{1}{\sqrt{\Ga_i(\Ga_i+\Gb_i)}}. \eeq{1.11}

The global solution is then determined by the local solutions $V_i(x,t)$ by considering suitable transmission conditions (continuity of potential and current flux) at the space--time boundaries between phases, at which the disturbance splits into two waves, as demanded by the weak form of \eq{1.1}. 
%
In general, if the space--time interfaces are not suitably placed, the disturbances will branch giving rise to a complicated cascade of current lines that is difficult to analyze. 
However, FP--materials are those materials in which the space--time geometry is such that the characteristic lines form an orderly pattern, the field pattern \cite{Milton:2017:FP,Mattei:2017:FPW}, with a locally periodic behavior in space and time (though not necessarily with the same unit cell as the underlying dynamic material, as
in \fig{Geometries}c). In \fig{Geometries} we show three FP--materials: \fig{Geometries}a and \fig{Geometries}b represent space--time checkerboards (see \cite{Mattei:2017:FPW} for details), whereas \fig{Geometries}c depicts a space--time composite formed by rectangular inclusions of material 2 immersed in material 1 (see \cite{Milton:2017:FP} for details). Space--time checkerboards were also studied by Lurie and coworkers (see, e.g., \cite{Lurie:2009:MAW}) who discovered that with no impedance
mismatch at interfaces, the trajectories of characteristics can converge at long times, somewhat analogously to shocks in a non-linear medium.  
\begin{figure}[!ht]
	\includegraphics[width=\textwidth]{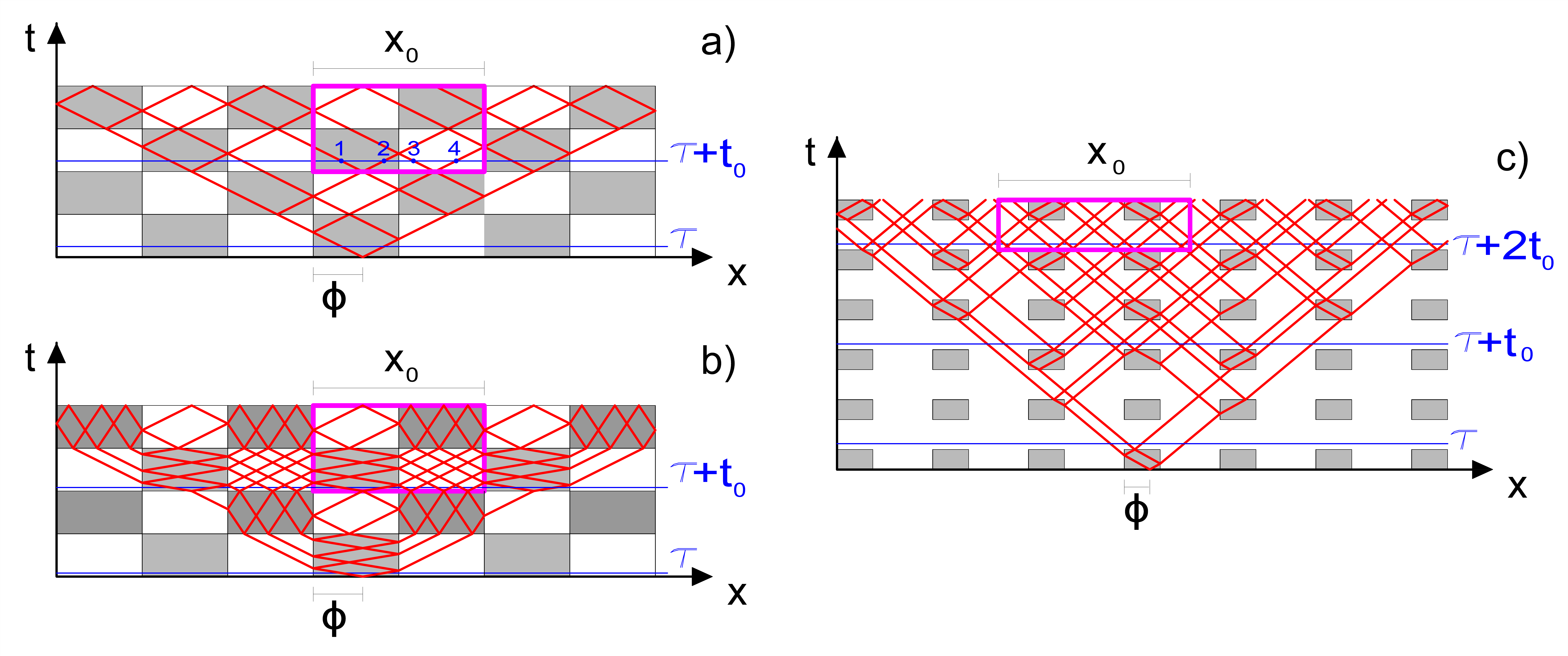}
	\caption{a) represents a two-phase space--time checkerboard in which the two phases have the same wave speed, so that the slope of the characteristic lines is the same in the two phases (see \cite{Mattei:2017:FPW}). The field pattern is shown (in red), and within each unit cell (in magenta) we calculate the current distribution only at four points, denoted by blue dots in the picture. $\phi$ is the launching parameter, that depends on the coordinate of the injection point $(a,0)$, and it parametrizes the field pattern. b) represents a three-phase space--time checkerboard (phase 1 is colored in white, phase 2 in light gray, and phase 3 in gray) in which the wave speeds of the phases satisfy the following relation: $c_2/c_1=c_1/c_3=3$ (see \cite{Mattei:2017:FPW}). c) represents a space--time microstructure in which rectangular inclusions of phase 2 are immersed in the space--time matrix made of material 1 (see \cite{Milton:2017:FP}). In this case the unit cell of the field pattern is twice that of the geometry. }
	\labfig{Geometries}     	
\end{figure}

\section{Dispersion diagrams}

As field patterns are periodic in space and time, we can use Bloch--Floquet theory to find solutions to the conductivity problem \eq{1.1}--\eq{1.7a}. In particular, we look for ``discrete solutions'' of the Bloch--Floquet type, as the discrete nature of field patterns allows one to determine the solution (in terms of ``electric currents'') just on the discrete network of characteristic lines. Specifically, due to the local time periodicity of field patterns, for a fixed value of the launching parameter $\phi$ (see \fig{Geometries}), we determine the distribution of currents along the characteristic lines at discrete moments of time, say $t=\tau+n\,t_0$ for $n=0,1,2,...$, where $t_0$ is the time period of the discrete network and $\tau$ is some fixed time (for simplicity $\tau$ is chosen in such a way that none of the characteristic lines
intersect, see \fig{Geometries}). Then, within each unit cell, the state of the system is captured by the function $j(l,m,n)$, where the integer $l$ indexes the current line within the unit cell ($l$ is the number of intersection points between the characteristic lines and the horizontal line $t=\tau$), the integer $m$ indexes the cell, and the integer $n$ indexes the discrete time (note that the unit cell of periodicity of the field pattern may be different from that of the microstructure, as in \fig{Geometries}c).


Now, according to the Bloch--Floquet theory, we suppose that (considering periodic boundary conditions)
\beq
j(l,m+s,n)=\mathrm{exp}(\mathrm{i}ks)\,j(l,m,n)\,,
\eeq{Bloch_space}  
that is, at each discrete moment of time ($n$ is fixed), the current at point $l$ of cell $m+p$ is proportional to the current at point $l$ of cell $m$ by the complex exponential $\mathrm{exp}(\mathrm{i}ks)$, where $k$ is the Bloch wavenumber and $s$ is an integer.

To determine the evolution of the state function $j(l,m,n)$ as $n=0,1,2,...$ increases, one has to calculate the Green function that allows one to recover the currents at a certain time $t=\tau+n\,t_0$ with $n$ fixed, given the currents at time $t=\tau+(n-1)\,t_0$. One does this by 
taking one unit cell, say with $m=m_0$ and injecting, at $t=\tau$ ($n=0$), a unit current in each of the $l$ intersection points, one at a time, and by calculating how such a current flows along the characteristic lines to determine the currents at $t=\tau+t_0$ ($n=1$) (note that the only material parameters that enter this calculation are the phases impedances $\gamma_i$, $i=1,\dots,p$, given by \eq{1.11}). We denote the Green function by $G_{l,l'}(m-m')$ to indicate that it provides the current at point $l$ of cell $m$, given the current at point $l'$ of cell $m'$. Then, 
\beq j(l,m,n)=\sum_{l',m'}T_{(l,m),(l',m')}j(l',m',n-1), \eeq{3.1}
where
\beq T_{(l,m),(l,'m')}=G_{l,l'}(m-m') \eeq{3.2}
is the transfer matrix, whose components depend only on the impedances $\gamma_i$ \eq{1.11} and, due to \eq{Bloch_space}, on the Bloch wavenumber $k$, since $j(l',m',n-1)=\mathrm{exp}(\mathrm{i}k(m'-m))\,j(l',m,n-1)$ (for the explicit expression of the transfer matrix for each microstructure in \fig{Geometries} we refer the reader to \cite{Milton:2017:FP,Mattei:2017:FPW}). Consequently, the eigenvalues $\lambda$ of the transfer matrix are functions of $k$, i.e., $\lambda=\lambda(k)$. Note that, within the same unit cell (fixed $m$), the distribution of current at $t=\tau+n\,t_0$ and that at $t=\tau+(n+q)\,t_0$ are related by
\beq
j(l,m,n+q)=\lambda^q \,j(l,m,n)
\eeq{}
On the other hand, according to the Bloch--Floquet theory one has
\beq
j(l,m,n+q)=\mathrm{exp}(\mathrm{i}\,\omega\, q) j(l,m,n),
\eeq{}
where $\omega$ is the crystal frequency, and by comparison we have (for simplicity, suppose that $q=1$)
\beq \lambda(k)=\mathrm{exp}(\mathrm{i}\,\omega),
\eeq{dispersion}
that is the general expression for the dispersion relation of FP--materials. 

For the two--phase checkerboard geometry of \fig{Geometries}a the eigenvalues $\lambda(k)$ of the transfer matrix can be obtained explicitly.
We have $\lambda(k)=\mathrm{exp}(\pm \mathrm{i}k)$, so that the dispersion relation \eq{dispersion} turns trivially into $\omega=\pm k$ and the group velocity $\partial\omega/\partial k$ is simply equal to $\pm 1$, see the diagram on the left of \fig{check_2p_disp}. Note that $\omega$ does not depend on the choice of the impedances of the two phases, $\gamma_1$ and $\gamma_2$, and is real in accordance with numerical results \cite{Mattei:2017:FPW} that show the transfer matrix 
has all its eigenvalues on the unit circle (the case of unbroken PT--symmetry) for any choice of $\gamma_1$ and $\gamma_2$, so that all the modes are propagating modes. 
In particular, the dispersion relation is exactly the same as when $\gamma_1=\gamma_2$, corresponding to the trivial case of a homogeneous material. Significantly, however, the Green 
function is radically different. An instantaneous disturbance generated by a unit current injected at $t=0$ (in \fig{check_2p_disp}, right, this corresponds to $n=0$)
has fronts that travel at the group velocity  (in \fig{check_2p_disp}, right, the slope is equal to $1/4$ which corresponds to a group velocity $\pm 1$ as the number of current lines within each unit cell is equal to $l=4$). In contrast
to the case when $\gamma_1=\gamma_2$, the fronts are trailed by an oscillating wave with an amplitude (dependent on $\gamma_1$ and $\gamma_2$) that does not tend to zero as one moves away from the front:
this is an entirely new class of wave that we call a field pattern wave (FP--wave), see also
Figure 16 in \cite{Mattei:2017:FPW} for a different example of a FP--wave. This example shows that the band structure alone is insufficient to determine the long-time response to a localized disturbance. 

\begin{figure}[!ht]
	\includegraphics[width=\textwidth]{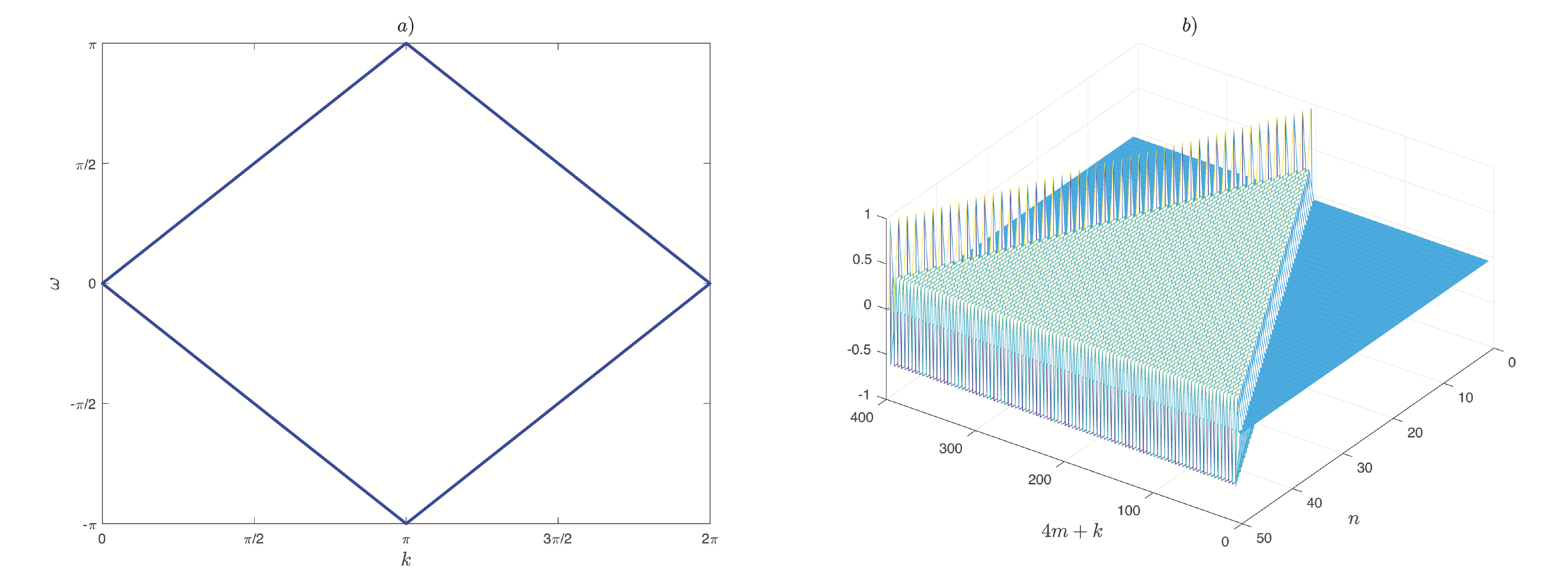}
	\caption{On the left is the dispersion diagram of the two-phase space--time checkerboard of \fig{Geometries}a. The dispersion relation is independent of the impedances $\gamma_1$ and $\gamma_2$. On the right is an example of a FP--wave propagating through the space--time checkerboard of \fig{Geometries}a, with $\Gg_1=1$ and $\Gg_2=4$, when 100 unit cells are considered (since there are 4 points to monitor in each unit cell, the total number of points is equal to 400). We inject a unit current at point 1 of cell 50, that is, at the point labeled with 201, and as time evolves, that is, for $t=\tau+nt_0$ with $n=0,1,\dots,50$, the current flows with the same intensity at the edge of a triangular shape having vertex at the point of injection of current. Inside the triangular shape the current oscillates and has zero average value.}
	\labfig{check_2p_disp}     	
\end{figure}


For the three--phase space--time microstructure of \fig{Geometries}b, the dispersion diagrams are shown in \fig{check3_disp} for different combinations of the phases impedances. We recall from \cite{Mattei:2017:FPW}, that the PT--symmetry condition for this microstructure is broken (leading to FP--waves with time exponential blow up or time exponential decay corresponding to complex $\omega$) only when either $\gamma_3\leq\gamma_1$ and $\gamma_3\leq\gamma_2$, or when $\gamma_1\leq\gamma_2\leq\gamma_3$. For all the remaining combinations, the PT--symmetry condition is unbroken (the field pattern supports only propagating modes and $\omega$ is real). The transition between a condition of unbroken PT--symmetry to a condition of broken PT--symmetry corresponds to the case when $\gamma_2=\gamma_3$ for which the dispersion diagram, shown in \fig{check3_disp}c, coincides with that depicted in \fig{check_2p_disp}a, corresponding to the two--phase space--time checkerboard of \fig{Geometries}a. Note that complex values of $\omega$, and hence blowup, occur when the number of real values of $\omega$ (between $-\pi$ and $\pi$) at a given $k$ changes with $k$ and, when $\omega$ is complex, its real part is on a plateau.

\begin{figure}[!ht]
	\includegraphics[width=\textwidth]{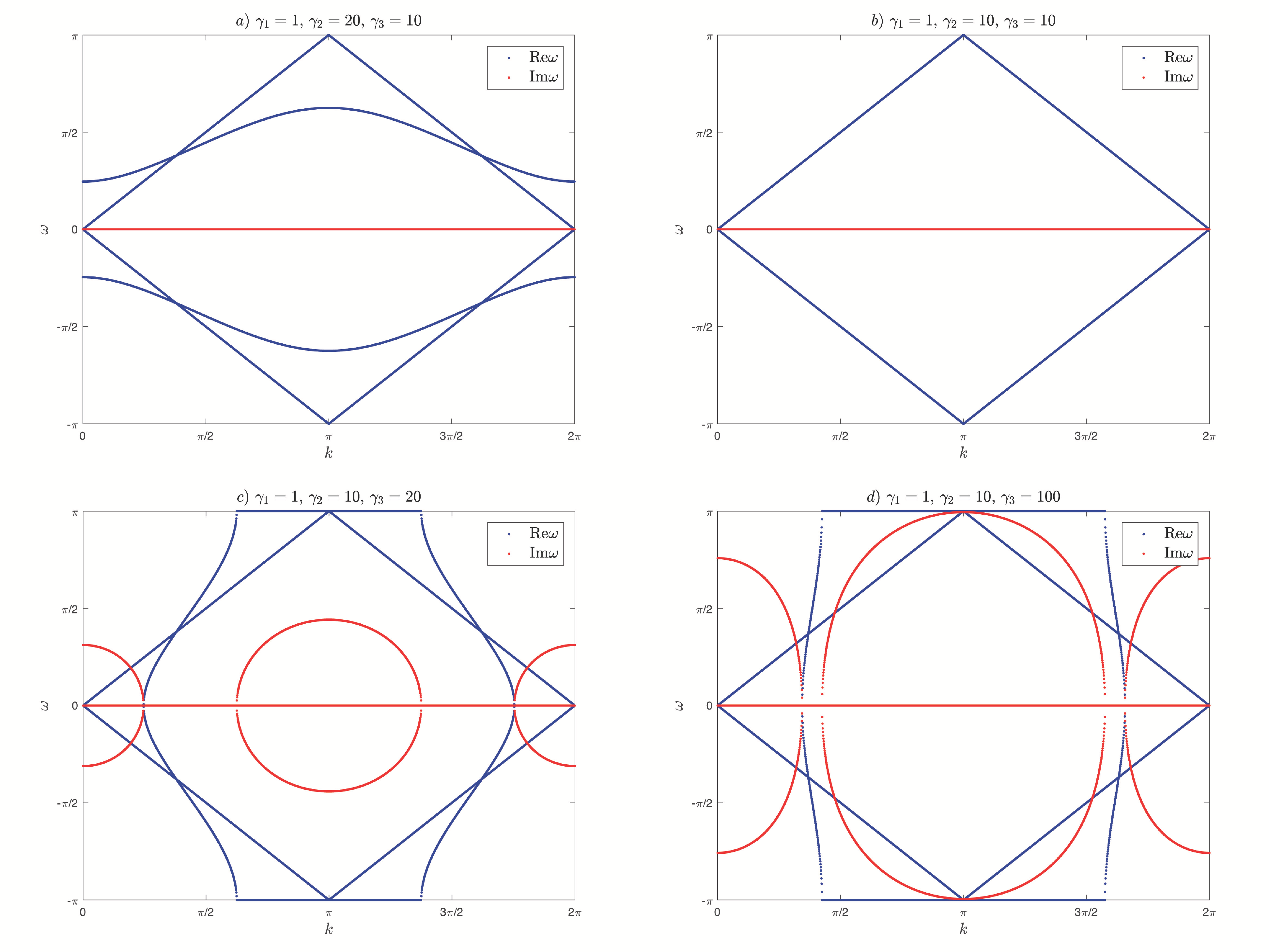}
	\caption{Dispersion diagrams for the three-phase space--time checkerboard shown in \fig{Geometries}b for different values of the impedances $\gamma_i$, $i=1,2,3$. The dispersion diagram $a)$ corresponds to a condition of unbroken PT--symmetry: $\omega$ is real and the field pattern supports only propagating modes. The transition between a condition of unbroken PT--symmetry to a condition of broken PT--symmetry is represented by the case $\gamma_2=\gamma_3$, see $b)$. The diagrams in $c)$ and $d)$ correspond to the condition of broken PT--symmetry: for some
$k$, $\omega$ can be complex and the field pattern can support, besides propagating modes, also modes that blow up or decay exponentially with time.}
	\labfig{check3_disp}     	
\end{figure}

In the microstructure with aligned rectangular inclusions, the PT--symmetry is unbroken only for the trivial case of no impedance mismatch (see \cite{Milton:2017:FP}), i.e., $\gamma_1=\gamma_2$, (note that $c_1\neq c_2$, so this does not correspond to the homogeneous case), and accordingly, $\omega$ is real, as shown in \fig{aligned_disp}a. If the impedances do not match
the field pattern supports not only propagating modes but also modes that blow up exponentially with time (coupled with modes that decay exponentially with time). Accordingly, the crystal frequency $\omega$ becomes complex (as in the previous case, this occurs when the number of real values of $\omega$ at a given $k$ changes with $k$, and when this happens, the real part of $\omega$ is on a plateau). Band gaps occur and are bigger the higher is the ratio between the two impedances (\fig{aligned_disp} shows only the case where $\gamma_2$ is increasingly bigger than $\gamma_1$ but similar results also hold when $\gamma_2$ is increasingly smaller). 

\begin{figure}[!ht]
	\includegraphics[width=\textwidth]{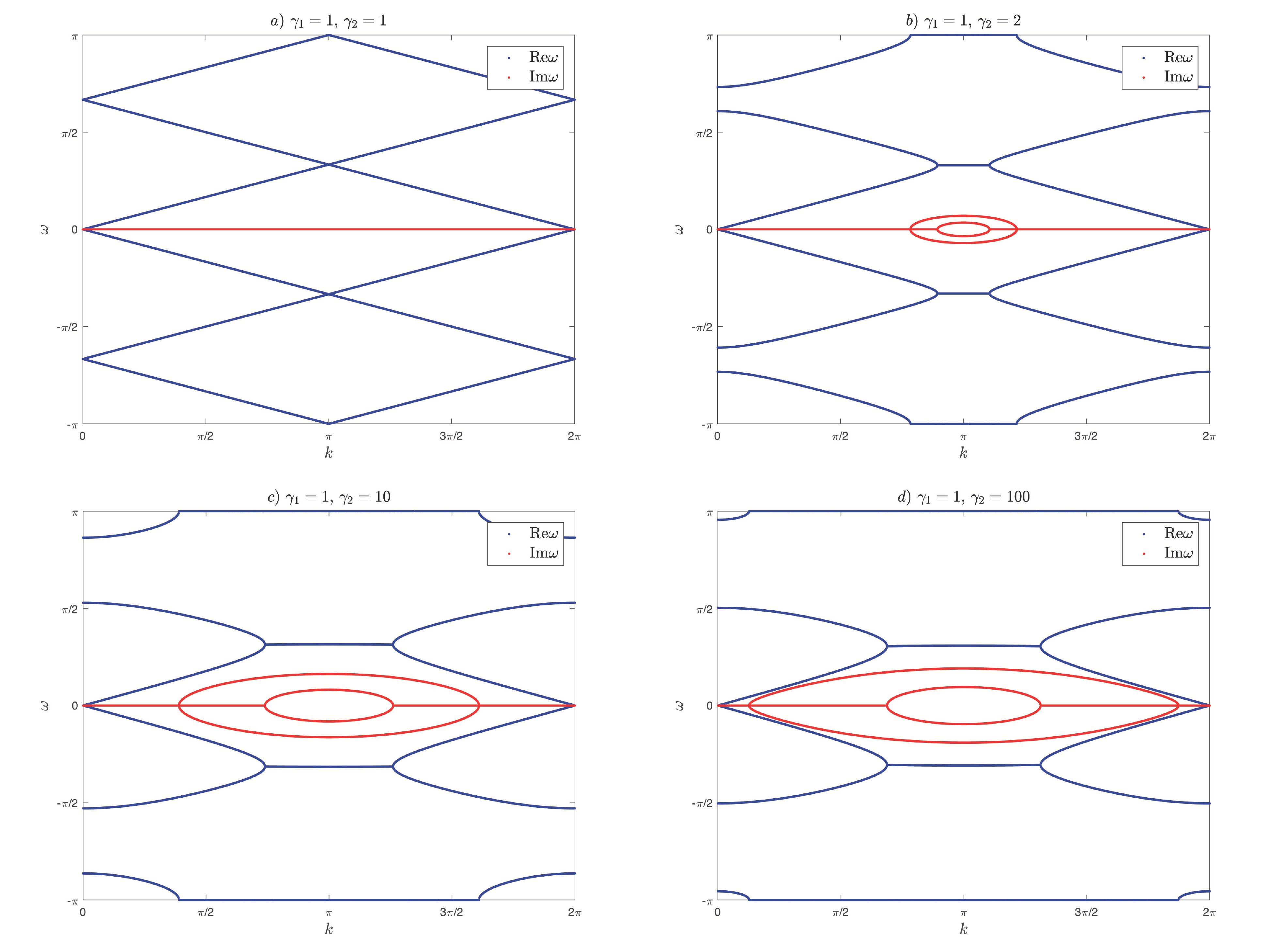}
	\caption{Dispersion diagrams for the space-time microstructures with aligned rectangular inclusions shown in \fig{Geometries}c for different values of the impedances $\gamma_1$ and $\gamma_2$. If there is no impedance mismatch, then the PT--symmetry condition is unbroken and $\omega$ is real. For any other choice of the values of the impedances, the PT--symmetry condition is broken (the field pattern supports not only propagating modes but also modes that blow up exponentially with time, together with modes that decay exponentially with time), and for some
$k$, $\omega$ can be complex. Furthermore, when $\gamma_1\neq\gamma_2$, band gaps appear.}
	\labfig{aligned_disp}     	
\end{figure}

The remarkable feature of these dispersion diagrams is that they are independent of the launching parameter $\phi$. This is what causes the infinite degeneracy of the band-structure.
Waves on different field patterns evolve entirely independently. An obvious interesting question, yet to be
explored, is how the band structure changes when the system is perturbed? One interesting possibility is that the perturbed spectrum will have an essential singularity. 
Essential singularities
are known to be associated with unusual phenomena such as anomalous resonance \cite{Nicorovici:1994:ODP} (that is the essential mechanism behind superlensing \cite{Pendry:2000:NRM}) 
and cloaking due to anomalous resonance (see \cite{Milton:2016:AM} and references therein).

\section*{Acknowledgments}
The authors are grateful to the University of Minnesota Institute for Mathematics and its Applications for support as part of
the special year on Mathematics and Optics. They also thank the National Science Foundation for support through
grant DMS-1211359. An anonymous scientist at a meeting is thanked for raising the question of what dispersion relations
field patterns have.

\ifx \bblindex \undefined \def \bblindex #1{} \fi\ifx \bbljournal \undefined
  \def \bbljournal #1{{\em #1}\index{#1@{\em #1}}} \fi\ifx \bblnumber
  \undefined \def \bblnumber #1{{\bf #1}} \fi\ifx \bblvolume \undefined \def
  \bblvolume #1{{\bf #1}} \fi\ifx \noopsort \undefined \def \noopsort #1{}
  \fi\ifx \bblindex \undefined \def \bblindex #1{} \fi\ifx \bbljournal
  \undefined \def \bbljournal #1{{\em #1}\index{#1@{\em #1}}} \fi\ifx
  \bblnumber \undefined \def \bblnumber #1{{\bf #1}} \fi\ifx \bblvolume
  \undefined \def \bblvolume #1{{\bf #1}} \fi\ifx \noopsort \undefined \def
  \noopsort #1{} \fi

\end{document}